\documentclass[twocolumn,showpacs,epsfig,prl]{revtex4}
\usepackage{graphicx}
\usepackage{amssymb}

\usepackage{color,soul}

\begin{document}

\draft
\title{Quantum Szilard Engine}
\author{Sang Wook Kim$^{1,2}$}
\author{Takahiro Sagawa$^{2}$}
\author{Simone De Liberato$^{2}$}
\author{Masahito Ueda$^{2}$}
\affiliation{$^1$Department of Physics Education, Pusan National University, Busan 609-735, Korea}
\affiliation{$^2$Department of Physics, University of Tokyo, Tokyo 113-0033, Japan}
\date{\today}

\begin{abstract}
The Szilard engine (SZE) is the quintessence of Maxwell's demon, which can extract the work from a heat bath by utilizing information. We present the first complete quantum analysis of the SZE, and derive an analytic expression of the quantum-mechanical work performed by a quantum SZE containing an arbitrary number of molecules, where it is crucial to regard the process of insertion or removal of a wall as a legitimate thermodynamic process. We find that more (less) work can be extracted from the bosonic (fermionic) SZE due to the indistinguishability of identical particles.
\end{abstract}
\pacs{03.67.-a,05.30.-d,89.70.Cf,05.70.-a}

\maketitle
\narrowtext

\begin{figure}
  \includegraphics[width=8cm]{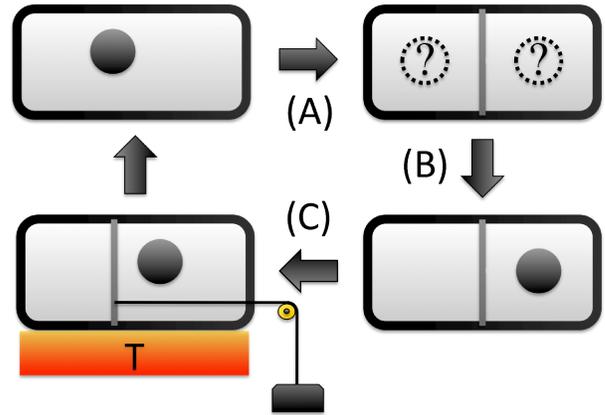}
  \caption{(Color online) Schematic diagram of the thermodynamic processes of the classical SZE. Initially a single molecule is prepared in an isolated box. (A) A wall depicted as a vertical gray bar is inserted to split the box into to two parts. The molecule is represented by the dotted circles to indicate that at this stage we do not know in which box the molecule is. (B) By the measurement, we find where the molecule is. (C) A load is attached to the wall to extract a work via an isothermal expansion at a constant temperature $T$.
  \label{fig1}}
\end{figure}

Maxwell's demon is a hypothetical being of intelligence that was conceived to illuminate possible limitations of the second law of thermodynamics \cite{Leff03,Maruyama09}. Leo Szilard conducted a classical analysis of the demon, considering an idealized heat engine with a one-molecule gas, and directly associated the information acquired by measurement with a physical entropy to save the second law \cite{Szilard29}. The basic working principle of the Szilard engine (SZE) is schematically illustrated in Fig.~\ref{fig1}. If one acquires the information concerning which side the molecule is in after dividing the box, the information can be utilized to extract work, e.g., via an isothermal expansion. The crucial question here is how this cyclic thermodynamic process is compatible with the second law; the entropy of $k_{\rm B} \ln 2$ ($k_{\rm B}$ is the Boltzmann constant) that the engine acquires during the isothermal process is not returned to the reservoir but seems to be accumulated inside the engine. Now it is widely accepted that the measurement process including erasure or reset of demon's memory requires the minimum energy cost of at least $k_{\rm B} \ln 2$, associated with the entropy decrease of the engine, and that it saves the second law \cite{Brillouin51,Landauer61,Bennett82,Sagawa09}.

Although the SZE deals with a microscopic object, namely an engine with a single molecule, its fully quantum analysis has not yet been conducted except for the measurement process \cite{Zurek84,Lloyd97}. In this Letter we present the first complete quantum analysis of the SZE. In the previous literature, it takes for granted that insertion or removal of the wall costs no energy. This assumption is justified in classical mechanics but not in quantum mechanics \cite{Bender05} because the insertion or removal of the wall alters the boundary condition that affects the eigenspectrum of the system. As shown below, a careful analysis of this process leads to a concise analytic expression of the total net work performed by the quantum SZE. If more than one particle is present in the SZE, we encounter the issue of indistinguishability of quantum identical particles. Indeed, how much work is extracted from the quantum SZE strongly depends crucially on whether it consists of either bosons or fermions. We also show that the crossover from indistinguishability to distinguishability gradually occurs as the temperature increases. We assume that the measurement is performed perfectly. The case of imperfect measurement is discussed in terms of mutual information in Ref.~\cite{Sagawa09}.

To define the thermodynamic work in quantum mechanics, let us consider a closed system described as $H\psi_n = E_n\psi_n$, where $H$, $\psi_n$ and $E_n$ are the Hamiltonian of the system, its $n$th eigenstate and eigenenergy, respectively. The internal energy $U$ of the system is given as $U = \sum_n E_n P_n$, where $P_n$ is the mean occupation number of the $n$th eigenstate. In equilibrium $P_n$ obeys the canonical distribution. From the derivative of $U$, one obtains $dU = \sum_n (E_n dP_n + P_n dE_n)$. Analogous to the classical thermodynamic first law, $TdS = dU + dW$, where $S$ and $W$ are the entropy and work done by the system, respectively, the quantum thermodynamic work (QTW) can be identified as $dW = -\sum_n P_n dE_n$ \cite{Kieu04,Esposito06}. Note that $\sum_n E_n dP_n$ should be associated with $TdS$ since the entropy $S$ is defined as $S = -k_{\rm B}\sum_n P_n \ln P_n$.

Although the process of inserting a wall is accompanied by neither heat nor work in the classical SZE, it is not the case with the quantum SZE. This process can be modeled as that of increasing the height of the potential barrier. In quantum mechanics, energy levels then vary, contributing to the QTW.
This process can be performed isothermally so that the temperature is kept constant during the whole process in conformity with the original spirit of the SZE. If the insertion is performed in an adiabatic process defined as $dQ = \sum_n E_n dP_n = 0$, one can easily show that the temperature is either changed or not well-defined at the end. The former is obvious considering the classical thermodynamic adiabatic process. If the wall is inserted in a quantum adiabatic manner, $dP_n=0$ is always satisfied. Given that the temperature is defined from the ratio of probabilities as $P_n/P_m = e^{-(E_n-E_m)/k_{\rm B}T}$, it is well-defined only if all energy differences are changed by the same ratio \cite{Quan07}. However, this cannot be achieved in the SZE because each energy level shifts in a different manner \cite{wall}.

To describe the quantum SZE it is indeed sufficient to know only the isothermal process for a whole cycle. If the external parameter $X$ is varied from $X_1$ to $X_2$ isothermally, the QTW is obtained 
as
\begin{eqnarray}
W &=& k_{\rm B}T \sum_n \int_{X_1}^{X_2} \frac{\partial \ln Z}{\partial E_n} \frac{\partial E_n}{\partial X} dX \label{eq:q_work_Z_int}\\
&=& k_{\rm B}T \left[\ln Z(X_2) - \ln Z(X_1)\right],
\label{eq:q_work_Z}
\end{eqnarray}
where $Z=\sum_n e^{-\beta E_n}$ is the partition function with $\beta = 1/k_{\rm B}T$.

It is worth mentioning that the isothermal process induces thermalization of the molecule with the reservoir at every moment, which destroys all the coherence among energy levels. Therefore, it is not necessary to describe the dynamics of our system in terms of the full density matrix; it is sufficient to know its diagonal part, i.e. the probabilities $P_n$. However, this thermalization has nothing to do with the measurement of the location of the molecule since it proceeds regardless of which box the molecule is in.

\begin{figure}
  \includegraphics[width=8.6cm]{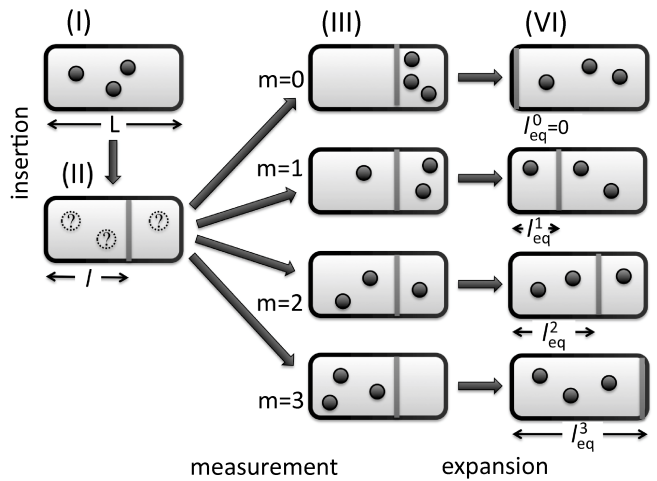}
  \caption{Schematic diagram of the quantum SZE containing three molecules. (I) Three molecules are prepared in a closed box with size $L$. (II) A wall, depicted by a vertical gray bar, is isothermally inserted at location $l$. The process (I) $\rightarrow$ (II) is called `insertion'. (III) The information on the number of molecules, $m$, on the left is acquired by the measurement. The process (II) $\rightarrow$ (III) is called 'measurement'. (IV) The wall moves and undergoes an isothermal expansion until it reaches its equilibrium location denoted by $l^m_{\rm eq}$. The process (III) $\rightarrow$ (IV) is called `expansion'. Finally the wall is isothermally removed to complete the cycle. The process (IV) $\rightarrow$ (I) is called `removal'.\label{fig2}}
\end{figure}

Let us now consider a quantum SZE in a general situation. As shown in Fig.~\ref{fig2}, the whole thermodynamic cycle consists of four processes, namely insertion, measurement, expansion and removal among four distinct states (I)-(IV). $N$ ideal identical molecules are prepared in a potential well of size $L$ as shown in Fig.~\ref{fig2}(I). A wall is then isothermally inserted at a certain position $l$. The partition function, at the moment when the wall insertion is completed but the measurement is not performed yet, is given as $Z(l) = \sum_{m=0}^N Z_m(l)$, where $Z_m(l)$ denotes the partition function for the case in which $m$ particles are on the left side and $N-m$ on the right. The amount of work required for the insertion process is thus expressed as
\begin{equation}
W_{\rm ins} = k_{\rm B}T \left[ \ln Z(l) - \ln Z(L) \right].
\label{eq:work_ins}
\end{equation}
Then, the measurement is performed without any expenditure of work. The amount of work extracted via the subsequent isothermal expansion is given as
$W_{\rm exp} = k_{\rm B}T \sum_{m=0}^N f_m\left[ \ln Z_m(l^m_{\rm eq}) - \ln Z_m(l) \right],$
where $f_m = Z_m(l)/Z(l)$ represents the probability of having $m$ particles on the left at the measurement. The wall moves until it reaches an equilibrium position $l^m_{\rm eq}$ determined by the force balance, $F^{\rm left}+F^{\rm right}=0$, where the generalized force $F$ is defined as $\sum_n P_n ({\partial E_n}/{\partial X})$, as illustrated in Fig.~\ref{fig2}(IV). We note that $l^{m}_{\rm eq}$ is not simply $(m/N)L$ unlike classical ideal gases.

The wall is then finally removed isothermally. In reality the wall is not impenetrable, and has a finite potential height, namely $X_\infty$. During the expansion process, $X_\infty$ is assumed to be large enough to satisfy $\tau_t \gg \tau$ , where $\tau_t$ and $\tau$ are a tunneling time between the two sides and an operational time of thermodynamic processes, respectively, ensuring that $m$ is well defined. During the wall removal, however, $\tau_t$ gradually decreases and becomes comparable with $\tau$ for a certain strength, $X_0$, where any eigenstate is delocalized over both sides due to tunneling. It implies that the partition function is given by $Z(l^m_{eq})=\sum_{n=0}^N  Z_n(l^m_{\rm eq})$ rather than $Z_m(l^m_{\rm eq})$. The integral (\ref{eq:q_work_Z_int}) for each $m$ can be split into two parts; $\int_{X_\infty}^{X_0} [\partial \ln Z_m(l^m_{\rm eq})/\partial X] dX$ and $\int_{X_0}^{0} [\partial \ln Z(l^m_{eq})/\partial X ]dX$. It is then shown that the former vanishes as far as the quasi-static process, $\tau \rightarrow \infty$ (i.e. $X_0, X_\infty \rightarrow \infty$), is concerned. This leads us to
\begin{equation}
W_{\rm rem} = k_{\rm B}T \sum_{m=0}^N f_m\left[ \ln Z(L) - \ln Z(l^m_{\rm eq}) \right].
\label{eq:W_rem}
\end{equation}
Note that the summation over $m$ must be made in $Z(l)$ of Eq.~(\ref{eq:work_ins}) for the insertion process irrespective of tunneling since no measurement has been performed yet.

Combining all the contributions the total work performed by the engine during a single cycle is given by
\begin{equation}
W_{\rm tot}= W_{\rm ins} + W_{\rm exp} + W_{\rm rem} = -k_{\rm B}T \sum_{m=0}^N f_m \ln \left(\frac{f_m}{f^*_m}\right),
\label{eq:W_tot}
\end{equation}
where $f^*_m = Z_m(l^m_{\rm eq})/Z(l^m_{\rm eq})$. Equation~(\ref{eq:W_tot}) has a clear information-theoretic interpretation in the context of the so-called relative entropy \cite{Vedral02} or Kullback-Leibler divergence even though $f^*_m$ is not normalized, namely $\sum_m f^*_m \neq 1$. It has been shown that the average disspative work upon bringing a system from one equilibrium state at a temperature $T$ into another one at the same temperature is given by the relative entropy of the phase space distributions between forward and  backward processes\cite{Kawai07}. With filtering or feedback control like measurement processes of the SZE that determine $m$, the work is represented as a form equivalent to Eq.~(\ref{eq:W_tot}) \cite{Parrondo09}.

\begin{table}
\caption{Total work measured in units of $k_{\rm B}T$ of the quantum SZE with $l=L/2$ containing two bosons or two fermions at the low and the high temperature limits (see \cite{suppl} for detailed derivation). \label{tab1}}
\begin{tabular}{c||c|c}
\hline
 & bosons & fermions \\ \hline
$T \rightarrow 0$ & $(2/3)\ln 3$ & $0$\\
$T \rightarrow \infty$ & $\ln 2$ & $\ln 2$ \\
\hline
\end{tabular}
\end{table}

It is straightforward to apply Eq.~(\ref{eq:W_tot}) to the original SZE consisting of a single molecule of mass $M$. For simplicity let us consider an infinite potential well of size $L$, and $l=L/2$. One finds $f^*_0=f^*_1=1$ since in these cases the wall reaches the end of the box so that $Z(l^m_{\rm eq}) = Z_m(l^m_{\rm eq})$ ($m=0, 1$) is satisfied. Note that $f^*_m=1$ is always true for $m=0$ and $N$. Together with $f_0=f_1=1/2$, we obtain $W_{\rm tot} = k_{\rm B}T \ln 2$, implying the work performed by the quantum SZE is equivalent to that of the classical SZE. However, consideration of individual processes reveals an important distinction between the classical and quantum SZE's. For the quantum SZE one obtains $W_{\rm ins}=-\Delta + k_{\rm B}T \ln 2$, $W_{\rm exp} = \Delta$ and $W_{\rm rem} =0$ for each process, where $\Delta = \ln [z(L)/z(L/2)]$, $z(l) = \sum_{n=1}^\infty e^{-\beta E_n(l)}$, and $E_n(l) = h^2n^2/(8 M l^2)$ with $h$ being the Planck constant. In the low-temperature limit, $\Delta$ is simply given as $E_1(L/2) - E_1(L)$. If the insertion process were ignored in the classical SZE, the second law would be violated because $\Delta \gg k_{\rm B}T$ in the low-temperature limit. In fact, $\Delta$ of the expansion process is compensated by the work required for inserting the wall. In the end, a tiny difference of work between these two processes results in the precise classical value, $k_{\rm B}T \ln 2$. As the temperature increases, the classical results of individual processes are recovered, i.e. $W_{\rm ins} \rightarrow 0$, $W_{\rm exp} \rightarrow k_{\rm B}T \ln 2$ and $W_{\rm rem} =0$ since $\Delta$ approaches $k_{\rm B}T \ln2$ in this limit.

For the quantum SZE with more than one particle dramatic quantum effects come into play. Let us consider a two-particle SZE confined in a symmetric potential well and a wall at $l=L/2$. One also finds $f^*_0 = f^*_2 = 1$ because of the same reason mentioned above. Since for $m=1$ the wall does not move in the expansion process, implying $l=l^1_{\rm eq}$, one obtains $f_1=f^*_1$. We thus end up with
\begin{equation}
W_{\rm tot}= - 2k_{\rm B}T f_0 \ln f_0,
\label{eq:W_two}
\end{equation}
where $f_0=f_2$ is used.

To get some physical insights let us consider two limiting cases of Eq.~(\ref{eq:W_two}), which is summarized in Table \ref{tab1} (see \cite{suppl} for a detailed derivation). For simplicity here the spin of a particle is ignored. In the low-temperature limit only the ground state is predominantly occupied, so that there exists effectively only one available state for each side. It is clear, as shown in Fig.~3(a), that for bosons $f_0$ should become $1/3$, i.e. $W_{\rm tot} = k_{\rm B}T (2/3) \ln2$, since we consider two mutually indistinguishable bosons over two places. On the other hand, two fermions are prohibited to be in the same side or to occupy the same state due to the Pauli exclusion principle, which explains why the work vanishes in the low-temperature limit. However, the higher the temperature, the larger the number of available states. Thermal fluctuations wash out indistinguishability since two identical particles start to be distinguished by occupying different states. This is why in the high-temperature limit we have $f_0=1/4$, i.e. $W_{\rm tot} = k_{\rm B}T \ln2$, for both bosons and fermions, which results from allocating two distinguishable particles over two places as shown in Fig.~3(b). It is also shown in Fig.~3 that $f_0$ continuously varies from $1/3$ ($0$) to $1/4$, exhibiting the crossover from indistinguishability to distinguishability for bosons (fermions) as the temperature increases.

\begin{figure}
\includegraphics[width=8.6cm]{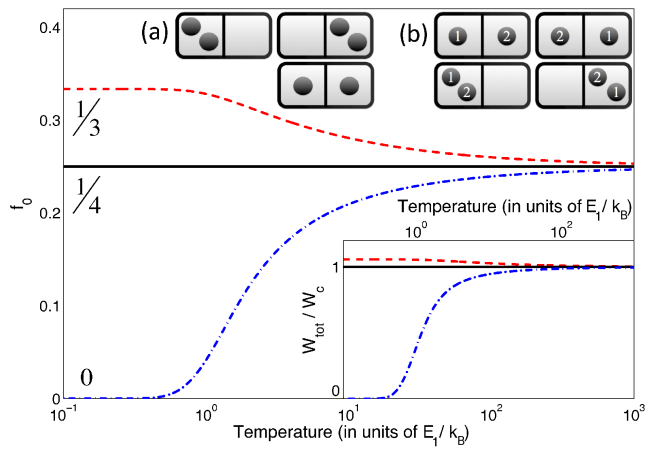}
  \caption{(Color online) $f_0$ as a function of $T$ for bosons (solid curve), fermions (dashed curve) and classical particles (dash-dotted line) in the case of the infinite potential well. The temperature is given in units of $E_1(L)/k_{\rm B}$. (a) Three possible ways in which two identical bosons are assigned over two states. (b) Four possible ways in which two distinguishable particles are allocated over two places. The inset shows $W_{\rm tot}/W_c$ as a function of $T$. \label{fig3}}
\end{figure}

The inset of Fig.~\ref{fig3} clearly shows that one can extract more work from the bosonic SZE but less work from the fermionic SZE over the entire range of temperature. (See \cite{suppl} for detailed discussions of the $W_{\rm tot}(T)$.) While details of $W_{\rm tot}(T)$ depend on the confinement potential, its low-temperature limits given in Table~\ref{tab1} are universal and have a deep physical meaning associated with the information content of quantum indistinguishable particles as mentioned above.

Finally, we briefly mention possible experimental realizations of the quantum SZE. Although several experiments \cite{Serreli07,Price08,Thorn08,Toyobe10} or proposals \cite{Scully03,Kim05} associated with the classical SZE have been presented so far, its fully quantum version has been elusive. There exist three important ingredients for the experimental realization: (i) controllability of the confinement potential, (ii) availability of a thermal heat bath to perform isothermal processes, and (iii) measurability of the work performed. For bosons, the system of trapped cold atoms may be a good candidate since the confinement potential can be easily controlled. Although such a system usually lacks a thermal heat bath, it can be immersed in a different species of atom trapped in a wider confinement potential so that they can play a role of a heat reservoir. For fermions, two-dimensional electron gases confined in quantum dots made of semiconductor heterostructures might be a candidate due to its controllability of the confinement potential and the existence of a heat reservoir of the Fermi sea of electrons. In principle, the work is determined once both $E_n$ and $P_n$ are known during the entire isothermal processes for both bosons and fermions.

In summary, we have studied the quantum nature of the SZE. The total work performed by the quantum SZE is expressed as a simple analytic formula which is directly associated with the relative entropy in the classical limit. To correctly describe the quantum SZE the processes of inserting or removing the wall should be regarded as a relevant thermodynamic procedure. The quantum SZE consisting of more than one particle clearly shows the quantum nature of indistinguishable identical particles. We believe our finings shed light on the subtle role of information in quantum physics.

We would like to thank Takuya Kanazawa and Juan Parrondo for useful discussions. SWK acknowledges JSPS fellowship for supporting his stay in University of Tokyo. SWK was supported by the NRF grant funded by the Korea government (MEST) (No.2009-0084606 and No.2009-0087261). TS acknowledges JSPS Research Fellowships for Young Scientists (Grant No.208038). SDL acknowledges FY2009 JSPS Postdoctoral Fellowship for Foreign Researchers. MU was supported by a Grant-in-Aid for Scientific Research (Grant No.22340114) and the Photon Frontier Network Program of the Ministry of Education, Culture, Sports, Science and Technology, Japan.


\begin{thebibliography}{150}
\bibitem{wall}
For example, if the wall is located at the node of the wavefunction, the corresponding energy level never changes as the height of the wall increases even though other levels move substanially.
\end{thebibliography}
\end{document}


\title{EPAP: Quantum Szilard Engine}
\author{Sang Wook Kim$^{1,2}$}
\author{Takahiro Sagawa$^{2}$}
\author{Simone De Liberato$^{2}$}
\author{Masahito Ueda$^{2}$}
\affiliation{$^1$Department of Physics Education, Pusan National University, Busan 609-735, Korea}
\affiliation{$^2$Department of Physics, University of Tokyo, Tokyo 113-0033, Japan}
\date{\today}

\begin{abstract}
We present the derivation of Table~I and the detailed discussion of $W_{\rm tot}(T)$ in the high temperature limit.
\end{abstract}
\pacs{}
\maketitle
\narrowtext


\section{Derivation of Table I}

Even though $W_{\rm tot}$ obtained from Eq.~(6) depends on what the confinement potential is, in the limits $T \rightarrow 0$ and $T \rightarrow \infty$, it has a universal value under quite general conditions discussed below. We first consider the two-boson case for an arbitrary confinement potential $V(x)$ with reflection symmetry, $V(x)=V(-x)$. The wall is inserted at $x=0$, i.e. $l=L/2$ separating $V$ into two single-wells. The probability that two bosons are in the right well is given as
\begin{equation}
f_0 = \frac{d+1}{4d+2},
\label{eq:f0boson}
\end{equation}
where
\begin{equation}
d \equiv \frac{z(\beta)^2}{z(2 \beta)}
\label{k}
\end{equation}
with $z(\beta)$ is the partition function of a single particle in the single-well at temperature $T \equiv (k_{\rm B}\beta)^{-1}$.

To obtain universal values of $f_0$ in the high- and low-temperature limits, the key observation is that
\begin{equation}
d \to \infty \ (T \to \infty)
\label{high}
\end{equation}
and
\begin{equation}
d \to 1 \ (T \to 0)
\label{low}
\end{equation}
hold only with the following two assumptions:
\begin{enumerate}
\item The number of energy levels is infinite.
\item The single-particle ground state is not degenerate.
\end{enumerate}
We first prove Eq.~(\ref{high}). Since $z(\beta) > z(2\beta)$ holds, we have $d > z(\beta)$. On the other hand, we can show  that $z(\beta) \to \infty$ holds with $T \to \infty$ from assumption 1. Therefore we obtain Eq.~(\ref{high}). Next, from assumption 2, we have $d = (1 + 2e^{-\beta (E_1 - E_0)} + \cdots ) / (1 + e^{-2\beta (E_1 - E_0)} + \cdots )$, where $E_n$ is the $n$th energy level of the single-well. It immediately leads us to Eq.~(\ref{low}). From Eqs.~(\ref{high}) and (\ref{low}), we obtain
\begin{equation}
f_0 \to 1/4  \ (T \to \infty)
\end{equation}
and
\begin{equation}
f_0 \to 1/3 \ (T \to 0),
\end{equation}
which are potential-independent universal results and describe the distinguishability-indistinguishability crossover by varying temperature.
The total work is then given as
\begin{equation}
W_{\rm tot} \to k_{\rm B}T \ln 2 \ (T \to \infty)
\end{equation}
and
\begin{equation}
W_{\rm tot} \to (2/3)k_{\rm B}T \ln 3 \ (T \to 0),
\end{equation}
which are also model-independent.

A similar approach can be done for the spinless two-fermion case. $f_0$ is given as
\begin{equation}
f_0 = \frac{d-1}{4d-2},
\label{eq:f0fermion}
\end{equation}
where $d$ is also defined as (\ref{k}). After exactly the same assumptions and similar procedure done above one reaches
\begin{equation}
f_0 \to 1/4  \ (T \to \infty)
\end{equation}
and
\begin{equation}
f_0 \to 0 \ (T \to 0).
\end{equation}
Therefore, we have
\begin{equation}
W_{\rm tot} \to k_{\rm B}T \ln 2 \ (T \to \infty)
\end{equation}
and
\begin{equation}
W_{\rm tot} \to 0 \ (T \to 0).
\end{equation}

\begin{figure}
   \includegraphics[width=15cm]{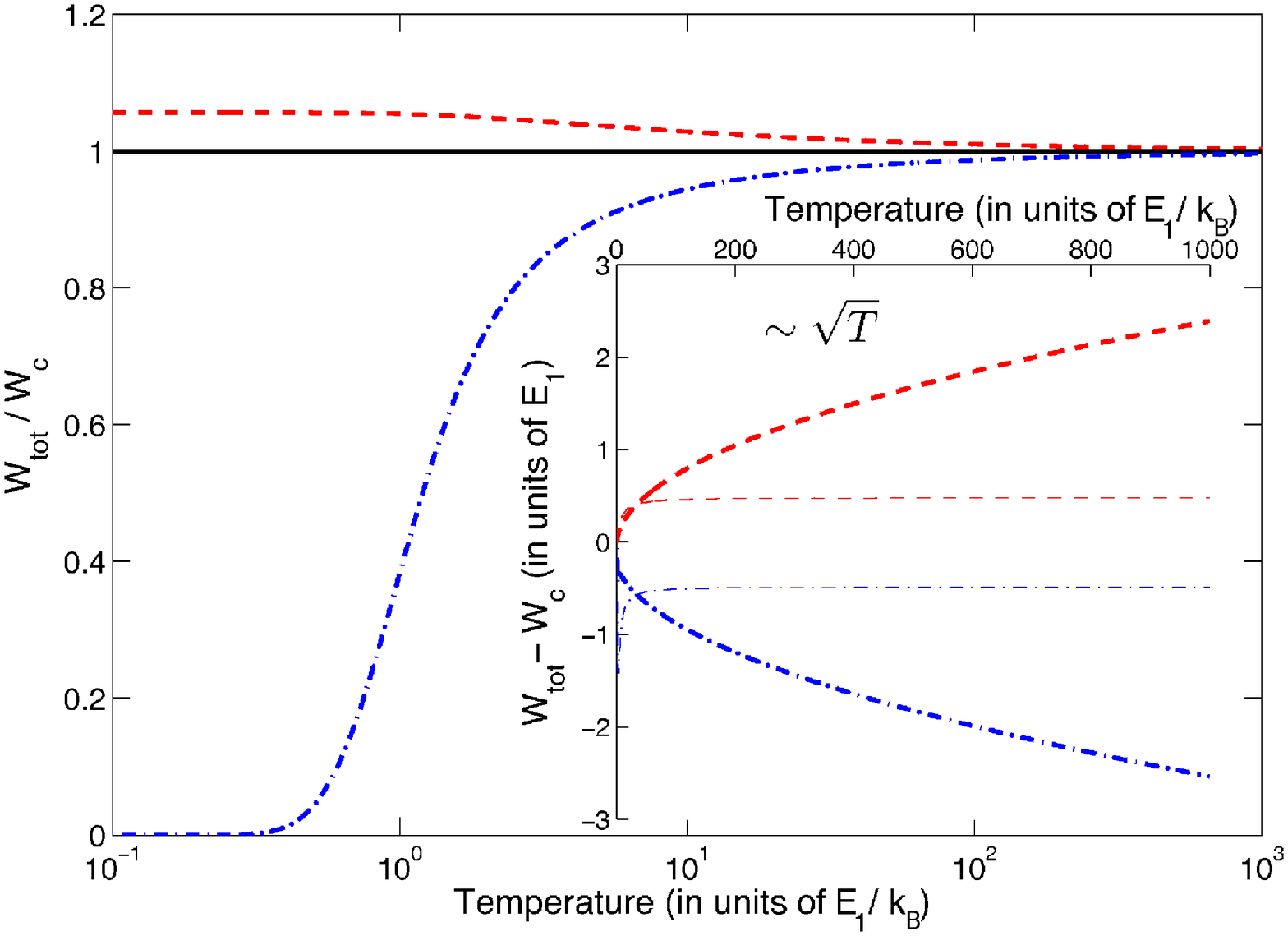}
  \caption{$W_{\rm tot}/W_c$ as a function of $T$ for bosons (solid curve) and fermions (dashed curve) obtained from Eq.~(5). The dashed dot line represents the classical work. This is the same as the inset of Fig.~3. The inset shows $W_{\rm tot}-W_c$ for bosons (solid curve) and fermions (dashed curve) as a function of $T$ for an infinite potential well (thick curves) and for a harmonic potential (thin curves), which exhibits the deviation proportional to $\pm T^{1/2}$ for the infinite potential well. (See the text for details.) For the harmonic potential $\hbar \omega = 10E_1$ is used, where $\omega$ and $E_1$ represent the resonant angular frequency of the harmonic oscillator and the energy of the ground state of the infinite potential well, respectively, and $\hbar=h/(2\pi)$. \label{fig4}}
\end{figure}

\section{$W_{\rm tot}$ as a function of $T$}

Figure \ref{fig4} shows $W_{\rm tot}/W_c$ is greater than one for bosons and smaller than one for fermions. As the temperature increases, $W_{\rm tot}/W_c$ approaches one for both bosons and fermions as expected. It is noted, however, that $W_{\rm tot}$ itself does not converge to $W_c$ but exhibits rather clear deviations from $W_c$ depending on the shape of the potential as shown in the inset of Fig.~\ref{fig4}. More precisely $W_{\rm tot}-W_c$ diverges as $\pm T^{1/2}$ ($+$ for bosons, and $-$ for fermions) for an infinite potential well, and keeps constant for a harmonic potential.

These deviations can be understood by expanding $W_{\rm tot}$ perturbatively. From Eqs.~(\ref{eq:f0boson}) and (\ref{eq:f0fermion}) one obtains
\begin{equation}
f_0=\frac{1 \pm b}{4 \pm 2b},
\end{equation}
where $b \equiv d^{-1}$ of Eq.~(\ref{k}), and $+$ for boson and $-$ for fermions. Note that $b$ goes to zero, i.e. $f_0 \rightarrow 1/4$, in the high temperature limit. The total work is then expressed as
\begin{equation}
W_{\rm tot} \approx k_{\rm B}T \left( \ln 2 \pm \frac{b}{4} \ln \frac{4}{e} \right).
\end{equation}
It is shown that $b \rightarrow T^{-1/2}$ for an infinite potential well implying $W_{\rm tot}-W_c \sim \pm T^{1/2}$, and $b \rightarrow T^{-1}$ for a harmonic potential implying $W_{\rm tot}-W_c \sim (k/4) \ln (4/e)$. In fact, one finds that $b$ approaches $T^{-1/\alpha}$ if the energy eigenvalues are given as $E_n \sim n^\alpha$. For $\alpha \rightarrow 0$, $W_{\rm tot}$ approaches the classical result irrespective of temperature. Since the infinite potential well is the steepest potential, $\alpha=2$ is the maximum possible value in practice. Therefore, at worst $W_{\rm tot}-W_c$ exhibits $\pm T^{1/2}$ deviation. Such a deviation is regarded as quantum effect of the SZE, and may be directly observed in experiment. Note that $b$ was ignored in the previous section, so that we simply had $W_{\rm tot} \rightarrow k_{\rm B}T \ln 2$ as $T \rightarrow \infty$ in Table~I.
